\documentclass[%
 aip,
 jcp, 
 nofootinbib,
 amsmath,amssymb,
 reprint,
]{revtex4-1}

\usepackage{graphicx}
\usepackage{dcolumn}
\usepackage{xcolor}
\usepackage{adjustbox}
\usepackage{import}
\usepackage{booktabs}
\usepackage{mathtools}
\usepackage{comment}
\usepackage{braket}
\usepackage{natbib}
\usepackage{siunitx}   

\usepackage{bm}

\usepackage[utf8]{inputenc}
\usepackage[T1]{fontenc}
\usepackage{mathptmx}
\usepackage{amsmath}
\usepackage{etoolbox}
\usepackage{hyperref}

\DeclareUnicodeCharacter{2009}{-} 
\DeclareUnicodeCharacter{2002}{\hspace{0.5em}}

\DeclareUnicodeCharacter{2500}{\textemdash} 
\DeclareUnicodeCharacter{2212}{-}

\makeatletter
\def\@email#1#2{%
 \endgroup
 \patchcmd{\titleblock@produce}
  {\frontmatter@RRAPformat}
  {\frontmatter@RRAPformat{\produce@RRAP{*#1\href{mailto:#2}{#2}}}\frontmatter@RRAPformat}
  {}{}
}%
\makeatother
\begin{document}

\preprint{AIP/123-QED}

\title
{
Selectively enabling linear combination of atomic orbital coefficients 
to improve linear method optimizations in variational Monte Carlo
}

\author{Trine Kay Quady${}^\dagger$}
 \affiliation{ 
Department of Chemistry, University of California, Berkeley, California 94720, USA
}

\author{Eric Neuscamman}
 \email{eneuscamman@berkeley.edu}

  \affiliation{ 
Department of Chemistry, University of California, Berkeley, California 94720, USA
}
\affiliation{%
Chemical Sciences Division, Lawrence Berkeley National Laboratory, Berkeley, California 94720, USA
}%

\date{\today}

\begin{abstract}
Second order stochastic optimization methods, such as the linear method,
couple the updates of different parameters and, in so doing, allow
statistical uncertainty in one parameter to affect the update of other
parameters.
In simple tests, we demonstrate that the presence of unimportant orbital
optimization parameters, even when initialized to zero, seriously degrade
the statistical quality of the linear method's update for important
orbital parameters.
To counteract this issue, we develop an expand-and-prune selective
linear combination of atomic orbitals algorithm that removes unimportant
parameters from the variational set on the fly.
In variational Monte Carlo orbital optimizations in propene, butene,
and pentadiene, we find that large fractions of the parameters can be
safely removed, and that doing so can increase the efficacy of the
overall optimization.
\end{abstract}

\maketitle

\section{\label{sec:intro}Introduction }

Optimization is central to the success of variational Monte Carlo (VMC). 
Provided an appropriate parameterization of the trial wave function, VMC is a flexible method that can produce high accuracy properties when optimized to the variational solution. 
Historically, VMC has been used either to
calculate ground-state energies,~\cite{mcmillanGroundStateLiquid1965,
ceperleyGroundStateElectron1980,
foulkesQuantumMonteCarlo2001,
luchowQuantumMonteCarlo2011,
austinQuantumMonteCarlo2012,
rubensteinIntroductionVariationalMonte2017} taking advantage of the zero-variance principle,~\cite{assarafZeroVariancePrincipleMonte1999}
or as the trial wave function in projector quantum Monte Carlo (QMC) methods, such as diffusion Monte Carlo.~\cite{reynoldsFixednodeQuantumMonte1982,
foulkesQuantumMonteCarlo2001,
needsContinuumVariationalDiffusion2009} 
Either alone or through providing a nodal surface, the application of VMC has grown to observables such as 
excited states,~\cite{shepardDoubleExcitationEnergies2022,
dashExcitedStatesSelected2019,
valssonPhotoisomerizationModelRetinal2010,
needsContinuumVariationalDiffusion2009}
magnetism and superconductivity,~\cite{busemeyerCompetingCollinearMagnetic2016,
heVariationalMonteCarlo2024,
sorellaPhaseDiagramHubbard2022}
and atomic forces~\cite{sorellaAlgorithmicDifferentiationCalculation2010,
valssonPhotoisomerizationModelRetinal2010,
nakanoEfficientCalculationUnbiased2024}
to name a few.
More recently, VMC has been adopted by the machine learning community, finding success with \textit{ab initio} neural networks as the VMC  ansatz.~\cite{carleoSolvingQuantumManybody2017,
pfauInitioSolutionManyelectron2020,
hermannDeepneuralnetworkSolutionElectronic2020,
hermannAbinitioQuantumChemistry2022}
Indeed, VMC has long been an effective tool for working
with ans\"atze
that lack efficient deterministic approaches, Jastrow factors being an important example.
However, the advantages offered by its statistical approach
come with challenges, in particular the challenges associated
with optimizing statistically uncertain objective functions.
By developing an approach to automatically screen out
physically unimportant parameters, specifically among
the molecular orbital coefficients, this study seeks to
improve VMC optimization by eliminating the uncertainty
these parameters introduce into other, more important
parameter's updates.

Like optimization algorithms more generally, optimizers for stochastic objective functions must make tradeoffs between the amount of information about the objective they employ and the cost, in time and memory, of obtaining and using that information.
First-order methods, such as steepest descent~\cite{beccaQuantumMonteCarlo2017,
sorellaWaveFunctionOptimization2005,
harjuStochasticGradientApproximation1997,
linOptimizationQuantumMonte2000} or accelerated descent,~\cite{sabzevariImprovedSpeedScaling2018,
reddiConvergenceAdam2019,
kingmaAdamMethodStochastic2017a,
schwarzProjectorQuantumMonte2017,
mahajanSymmetryProjectedJastrowMeanField2019}
widely used in the machine learning community,\cite{wang9603742,otisOptimizationStabilityExcitedStateSpecific2023}
require an amount of memory storage that grows only
linearly with the number of optimizable parameters $N_p$. 
These methods typically rely on using many iterations, each involving a minimal sampling effort, which results in very fast updates at the cost of long convergence tails.~\cite{otisComplementaryFirstSecond2019,
pengAnalysisFirstSecondOrder2025}
Comparatively, pseudo-second order methods such as the linear method (LM)
~\cite{toulouseFullOptimizationJastrow2008,
toulouseOptimizationQuantumMonte2007,
nightingaleOptimizationGroundExcitedState2001,
umrigarAlleviationFermionSignProblem2007}
and other quasi-Newton methods~\cite{umrigarEnergyVarianceOptimization2005}
incorporate some second-derivative information,
allowing coupling between parameters within the update. 
Second-order methods, with their quadratically scaling memory requirements
and their stronger dependence on sufficient sampling (see below),
have typically been limited to applications involving relatively small
sets of variational parameters.
\cite{pengAnalysisFirstSecondOrder2025,schwarzProjectorQuantumMonte2017}        
Attempts to bypass the memory bottleneck have included
the blocked LM, \cite{zhaoBlockedLinearMethod2017} 
hybridizations between the blocked LM and accelerated descent,
\cite{otisHybridApproachExcitedstatespecific2020}
and Krylov-subspace based solvers for second order update equations.
\cite{neuscammanOptimizingLargeParameter2012,sabzevariAcceleratedLinearMethod2020}
Although these approaches can reduce memory footprint, one also must
contend with the number of samples required to make second order updates stable.

In general, variational methods should improve as more and more flexibility
is added to the ansatz.
However, when flexibility is added by adding additional
variational parameters, the very coupling between
parameters' updates that is the strength of second order optimizers
can in fact lead to worse variational outcomes in practice
if statistical uncertainties in the couplings degrade the quality
of the update steps for the original parameters.
\cite{pengAnalysisFirstSecondOrder2025,
webberRayleighGaussNewtonOptimizationEnhanced2022} 
With sufficient sampling, this challenge can be overcome, and indeed in
this regime the LM has proven to be a more effective VMC optimizer than
its first order rivals.
\cite{otisOptimizationStabilityExcitedStateSpecific2023}
Likewise, as the ansatz flexibility approaches the point at which its
tangent space contains the exact wave function, its strong zero
variance principle allows it to mitigate this sampling concern.
\cite{pengAnalysisFirstSecondOrder2025,toulouseOptimizationQuantumMonte2007,nightingaleOptimizationGroundExcitedState2001} 
However, in many practical applications, neither of these regimes is easy
to reach, and including large numbers of variational parameters becomes
statistically challenging for the LM.
\cite{sabzevariAcceleratedLinearMethod2020,garnerImprovingVariationalMonte2023}
This issue is particularly pressing for 
the linear combination of atomic orbital (LCAO) coefficients used
to shape molecular orbitals (MOs), whose number grows quadratically with
system size.

Of course, in applications that support the use of localized MOs,
such as molecules and insulating materials,
it should be possible to mitigate this challenge by only allowing
$O(1)$ atomic orbitals to make nonzero contributions to a given MO,
at which point the LCAO parameters would grow only linearly with system size.
In the context of VMC, where there is no strict need for the MOs to be
orthogonal to one another, it should be possible to employ even more localized
orbitals than is possible in traditional quantum chemistry.
Here, we explore a local-orbital-based optimization algorithm
that automatically excludes unimportant LCAO parameters in order to
improve the statistics of the LM updates for those that are included
and, by doing so, enhance the overall efficacy of the optimization.
Through a series of tests on alkene molecules, we show that
this selected LCAO (sLCAO) approach can indeed deliver improved
LM optimization results, producing lower variational energies
than the standard LM using the same sampling effort.

\section{The Linear Method} 
\label{sec:LM}

The linear method involves a first order expansion of the normalized trial wave function 
\begin{equation} \label{norm}
 \overline{\Psi}(\mathbf{p}) = \frac{\Psi(\mathbf{p})}{\sqrt{\braket{\Psi(\mathbf{p})|\Psi(\mathbf{p})}}}
\end{equation}
in the self-plus-tangent space of the optimizable parameters, $p_i$,
\begin{equation} \label{lm_wfn}
\begin{split}
   \Psi(\mathbf{p}) = \Psi_0(\mathbf{p}) 
   + \sum_i^{N_p} \delta p_i \,\, \overline{\Psi^i}(\mathbf{p}) 
\end{split}
\end{equation}
where $N_p$ is the size of the variational parameter basis, $\Psi_0(\textbf{p})$ is the original trial wave function, and $\overline{\Psi^i}(\textbf{p})$ is the first order expansion of the wave function in the parameters \textbf{p} having been orthgonalized against $\Psi_0(\mathbf{p})$
\begin{equation} \label{lm_ddp}
\begin{split}
   \overline{\Psi^i} &= \frac{\partial \overline{\Psi}(\mathbf{p})}{\partial p_i} =  \frac{\partial \Psi(\mathbf{p})}{\partial p_i} - S_{0i} \, \Psi_0(\mathbf{p}) 
\end{split}
\end{equation}
where $S_{0i} = \braket{\Psi_0 |\Psi^i}$ and $\Psi^i=\partial\Psi/\partial p_i$.
The solution to the LM's parameter update, $\delta \mathbf{p}$, involves 
collecting the relevant terms 
$ \left( \textit{i.e.}  \frac{ \mathcal{H} \Psi }{\Psi}, \frac{\Psi^{i}}{\Psi}, \frac{\mathcal{H} \Psi^{i}}{\Psi} ... \right) $
during the Monte Carlo sampling procedure to construct the Hamiltonian matrix $\textbf{H}$ and overlap matrix $\textbf{S}$ in the 
LM's generalized eigenvalue problem within the $\{ \Psi_0, \overline{\Psi^i} \}$ basis.~\cite{toulouseOptimizationQuantumMonte2007,
toulouseFullOptimizationJastrow2008}
\begin{equation} \label{eq:Hdp=eSdp}
  \mathbf{H} \delta \mathbf{p} = \lambda \mathbf{S} \delta \mathbf{p} 
\end{equation}
\begin{equation} \label{HijandSij}
\begin{split}
   H_{ij} &= \left\langle \overline{\Psi^{i}} \Big| \mathcal{H} \Big| \overline{ \Psi^{j}} \right \rangle
   \quad \text{ and } \quad
   S_{ij} =  \left\langle \overline{ \Psi^{i}} \Big| \overline{\Psi^{j} } \right \rangle
\end{split}
\end{equation}
To stabilize the LM, a positive shift $a$ along the diagonal of the Hamiltonian is added: $H_{ij} \rightarrow H_{ij} + a\, \delta_{ij}(1-\delta_{i0})$, where $a = 0.01$ in all following results.
As the LCAO coefficients are nonlinear parameters, we also employ the
uniform rescaling of the update proposed by Toulouse and Umrigar,
\cite{toulouseOptimizationQuantumMonte2007}
in which the update is adjusted as
\begin{equation}
 \delta \mathbf{p} \leftarrow \frac{\delta \mathbf{p}}{1-\sum_i^{N_p} N_i \delta p_i}
\end{equation}
in which
\begin{equation}\label{eq:Ni_cyrus}
    N_i = \frac{(1-\zeta)\sum_j^{N_p} \delta p_j S_{ij}}{
    (1+\zeta)+\zeta \sqrt{1+\sum_{j,k}^{N_p} \delta p_j \delta p_k S_{jk}}
    }
\end{equation}
and $\zeta=1/2$.

\label{sec:unc-example}
\begin{figure*}[ht!]
    \centering
    \includegraphics[width=0.99\linewidth]{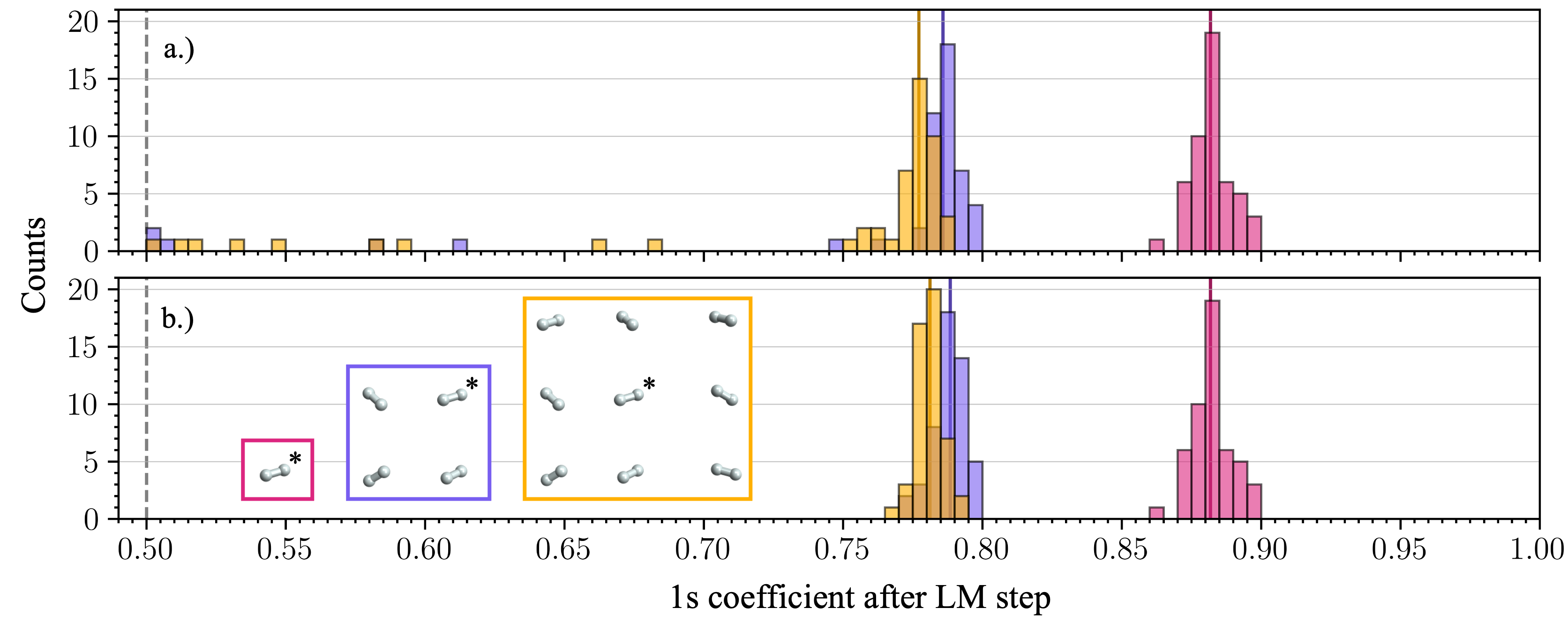}
    \caption{
    Histograms showing the updated LCAO coefficient for the
    1s orbital indicated by the asterisk within its molecule's MO
    in our step uncertainty test.
    The test was performed in three systems, which contained one (pink),
    four (purple), and nine (yellow) H$_2$ molecules (see SI for geometries).
    In panel (a), all LCAO coefficients are included as variational parameters.
    In panel (b), only the LCAO coefficients initially set to 0.5 are
    variational parameters (the asterisked 1s orbitals are in this set).
    Median updates are given by vertical lines.
    }
    \label{fig:H2study}
\end{figure*} 
\section{A step uncertainty example}
Before attempting to construct a method for automatically setting aside
LCAO coefficients that are not variationally helpful, we first look at
an example that demonstrates the problem.
In minimal-basis H$_2$, the $\sigma$ MO should contain equal
LCAO coefficients on each of the two 1s orbitals.
As an optimization test, we take collections of H$_2$ molecules
(either one, four, or nine of them) and intentionally skew the $\sigma$
MO within each molecule by setting one of its LCAO coefficients to 1.0
and the other to 0.5 (for a given molecule's MO, the coefficients on
other molecules' AOs are initialized to zero).
We then compare two types of LM updates, each of which we evaluated under
fifty different random seeds and with a sample size of 64,000 samples per H$_2$.
In the first case, shown in Figure \ref{fig:H2study}a,
\textit{all} of the LCAO coefficients are allowed to be variational parameters,
while in the second case, shown in Figure \ref{fig:H2study}b,
we restrict the set of variational parameters to those initially
set to 0.5, holding all others fixed.
In both cases, the median update of an initially-0.5 parameter
brought it most of the way towards equality with its initially-1.0 partner,
as chemical intuition would lead us to expect.
However, the distribution of updates for the 4H$_2$ and 9H$_2$ systems
contained a number of undesirable outliers when all of the LCAO parameters
were turned on, a problem that went away when the variable set was restricted
to the relatively small ``important'' set of initially-0.5 parameters.

\begin{figure}[]
    \centering
    \includegraphics[width=0.99\linewidth]{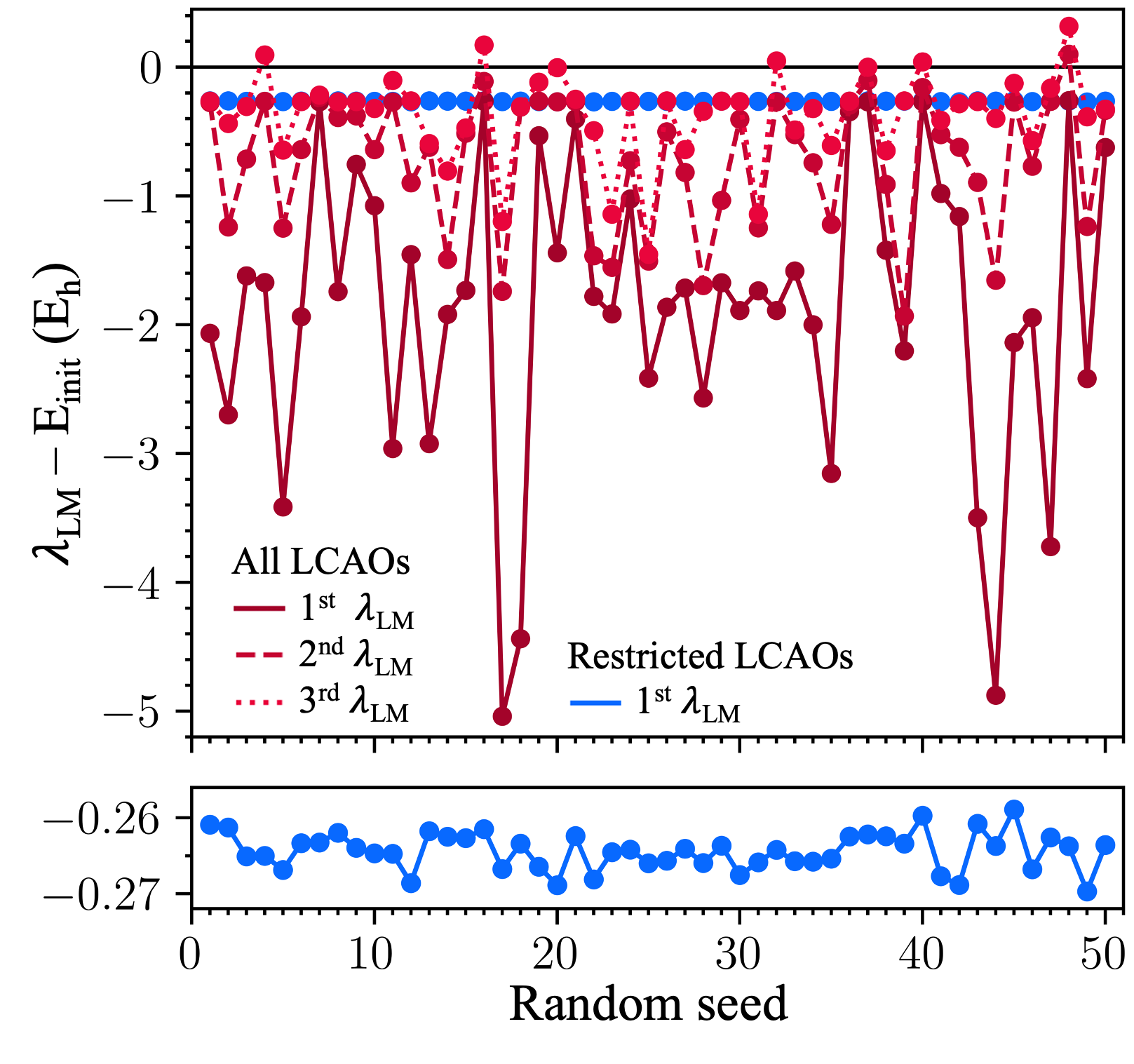}
    \caption{
    Predicted LM energy lowerings for the 9H$_2$ step uncertainty test,
    as given by the difference between the LM eigenvalue $\lambda$
    and the initial energy $\mathrm{E_{init}}$.
    The top panel shows the predictions corresponding to the lowest
    three eigenvalues for the all-LCAO case in red
    as well as that of the lowest eigenvalue for the reduced LCAO case in blue.
    The panel below provides a zoomed in view of the latter.
    } 
    \label{fig:H2eigenvals}
\end{figure}

To get an idea where these outliers were coming from, we analyzed the
spectrum of the LM eigenproblem.
Looking at the 9H$_2$ system, we see in  Figure \ref{fig:H2eigenvals}a
that, in the case where we include all LCAO coefficients as variational parameters,
the statistical uncertainty of the lowest three eigenvalues is large to the
point that the math within the LM tangent space is often predicting a multi-Hartree
improvement in the energy from this optimization step.
In fact, we know from a 
performing a complete 576,000
sample LM optimization on all LCAO coefficients
that the converged variational minimum in this setup
is only about 0.31 E$_\mathrm{h}$ below
the energy of our skewed-$\sigma$ initial guess.
The updates corresponding to the most extreme of the energy improvement predictions
were in fact so unphysical that, in order to get the mostly reasonable updates
displayed in Figure \ref{fig:H2study}a, we had to implement an energy update
threshold that selected the eigenvector corresponding to the lowest-energy
eigenvalue whose predicted energy lowering was not more than 0.3 E$_\mathrm{h}$.
Without this thresholding, the updates were straightforwardly unphysical.
In contrast, the LM eigenproblem produced much more reasonable predictions
of 0.26 to 0.27 E$_\mathrm{h}$ improvements when we limited the variational
set to the initially-0.5 coefficients, as seen in Figure \ref{fig:H2eigenvals}b.
Thus, although this is a contrived example, we clearly see that, when working
with a finite sample size, the statistical uncertainty introduced by adding
additional, less physically important LCAO parameters can in fact degrade the
quality of the optimization update.
To mitigate this issue, we now attempt to construct a selected LCAO scheme that
automatically filters out the physically unimportant and statistically
problematic parameters on the fly.

\label{sec:sLCAO}

\section{Expand-and-prune selected LCAO}

In constructing a selected LCAO (sLCAO) approach, we seek to strike a balance
between two competing priorities.
On the one hand, the more LCAO coefficients that are enabled within each
localized MO, the closer our ansatz will, at least in principle, be able to
come to its overall variational minimum.
On the other hand, the more coefficients that we can disable, the better
we expect the statistics of our LM updates to be.
Of course, in a sufficiently large system, we know that most LCAO coefficients
will be negligible for a localized orbital, so we should be able to find plenty
that can be safely disabled.
The question is how to produce a black-box compromise between our two
competing priorities.
Here, we strike the balance by performing a series of orbital expansions
and contractions during the overall LM optimization.
During each contraction, a single energetic threshold is used to prune
unimportant LCAO coefficients out of the variational parameter set.
Our anticipation is that, after many cycles of expansion and pruning,
the MOs will reach an equilibrium that establishes the
balance we seek.

Orbital pruning, which we apply both
to the initial Pipek-Mezey (PM)
\cite{pipekFastIntrinsicLocalization1989,lehtolaPipekMezeyOrbital2014}
localized RHF orbitals and during each LM update,
is based on a Fock-based estimate of each LCAO coefficient's
energetic importance.
Looking at each of the currently enabled coefficients, we
estimate that coefficient's importance via its contribution to
its MO's orbital energy, as estimated by the original RHF Fock operator $\hat{F}$.
For the $j$th MO, this orbital energy estimate is
\begin{equation}\label{E_F}
    \epsilon_j 
    = \frac{\Vec{c}^{{(j)}^\top} \mathbf{F}~\Vec{c}^{(j)}}
           {\Vec{c}^{{(j)}^\top} \mathbf{S}~\Vec{c}^{(j)} \rule{0mm}{4.2mm}}
\end{equation}
where
\textbf{F} and \textbf{S} are the atomic orbital Fock and overlap matrices,
respectively, and
\begin{align}
\bm{C} =
\begin{pmatrix}
\vline & \vline & \\
\Vec{c}^{(1)} \rule{0mm}{4.3mm} & \Vec{c}^{(2)} & \cdots \\
\vline & \vline & \\
\end{pmatrix}
\end{align}
is the LCAO matrix.
We then estimate the difference
\begin{equation}\label{eq:fock_le_ep}
 D_{ij} =  | \epsilon_j - \epsilon_{ij} |,
\end{equation}
in which $\epsilon_{ij}$ is the recalculation of $\epsilon_j$
after zeroing out $C_{ij}$.
The idea is to prune those coefficients that make little to no energetic
difference, and so our pruning step removes
from the variational parameter set all of the
LCAO coefficients for which $D_{ij}$ is less than a small threshold $\mu$.

\begin{figure}[]
    \centering
    \includegraphics[width=0.95\linewidth]{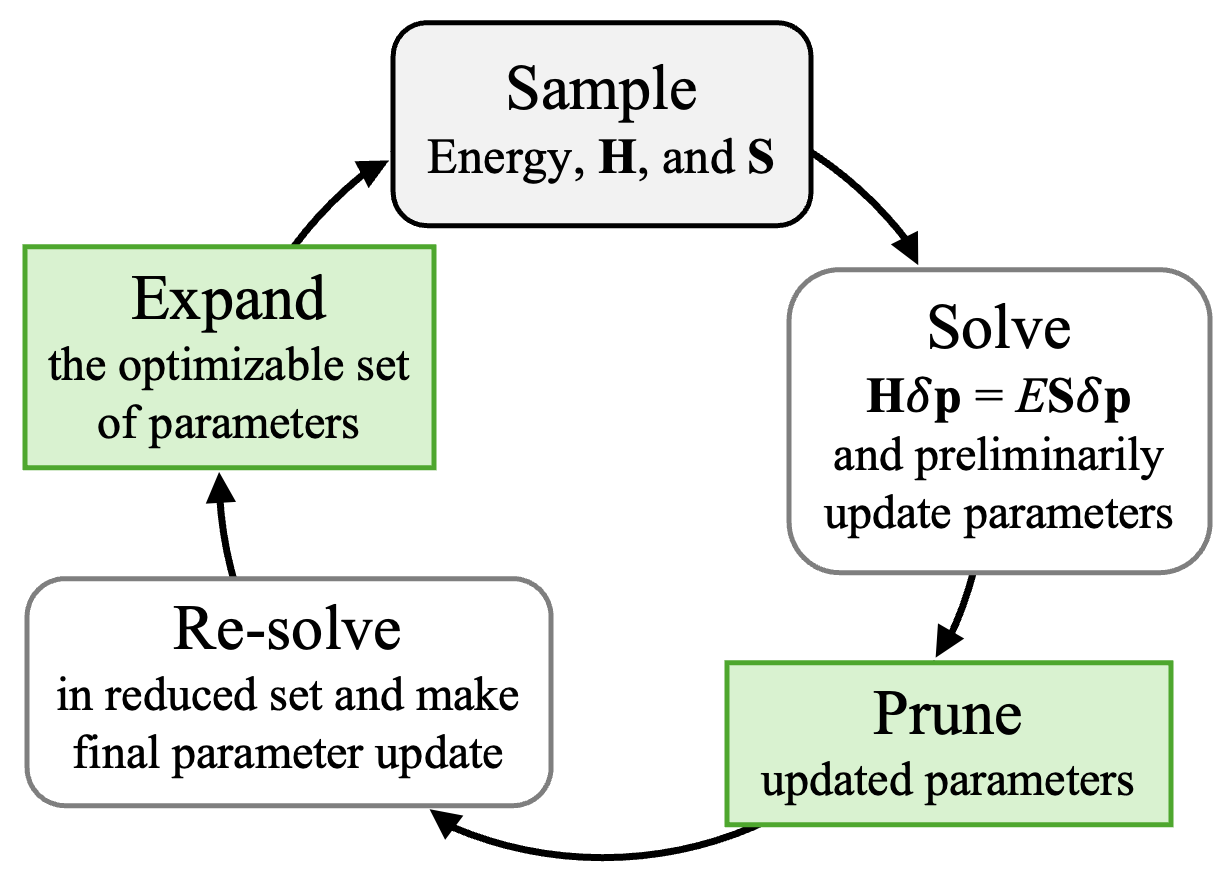}
    \caption{
    Schematic of our sLCAO method's cycle of orbital expansion and pruning.
    See Section \ref{sec:sLCAO} for details.
    }
    \label{fig:selection_cycle}
\end{figure}

As shown in Fig.\ \ref{fig:selection_cycle}, the sLCAO optimization proceeds
through a cycle of orbital expansions and prunings embedded within a modified LM.
To start, we (i) enable all LCAO coefficients that share an atom with one of the
coefficients that survived the initial pruning step.
We then (ii) take the VMC sample, construct the LM matrices
in the basis of the enabled variational parameters,
and solve for a ``suggested'' update for those parameters.
To minimize the noise in the update, we (iii) construct the MOs that this
suggested update would give us and prune them to arrive at a reduced set of
variational parameters.
In this reduced set, we (iv) re-solve the LM equations to produce
the actual parameter update, which we apply to produce the new wave function.
Finally, for each MO, we (v) enable any currently disabled LCAO
coefficients (initializing them to zero) that share an atom or are
one bond away from an atom on which that MO has an already-enabled coefficient.
At this point, we return to step (ii) and repeat the whole process,
producing an iteration in which the orbitals are allowed to expand as guided
by the variational principle but are kept reasonably local via the $\mu$-based pruning
of their energetically least important fringes.

\section{Computational  Details}

We employ a single-Slater Jastrow trial wave function
\begin{equation}
    \Psi_0(\textbf{X},\textbf{C}) = e^{J(\textbf{X})} \Phi (\textbf{X},\textbf{C})
\end{equation}
where $\mathbf{X}$ are the electron positions in real space and  
$\mathbf{C}$ are the LCAO coefficients.
$\Phi$ is a restricted Slater determinant,
which we initialize from an RHF calculation and Pipek-Mezey localization,
\cite{pipekFastIntrinsicLocalization1989,lehtolaPipekMezeyOrbital2014} 
both of which are performed using \textsc{Pyscf}.
\cite{sunLibcintEfficientGeneral2015,sunPySCFPythonbasedSimulations2018,sunRecentDevelopmentsPySCF2020}  
The electron-electron cusps are satisfied
via $J(\textbf{X})$, a simple single-parameter 
$(A)$ two-body Jastrow factor
\begin{equation}
 J(\textbf{X}; A) = \sum_{i,j}^{N} \frac{A}{r_{ij}} \bigg( 1-e^{\frac{-r_{ij}}{\sqrt{A \eta}}} \bigg)
\end{equation}
where $N$ is the total number of electrons, $r_{ij}$ is the distance 
between the $i$ and $j$th electrons in $\mathbf{X}$ and  $\eta = 2 \text{ or } 1$ 
for $\sigma_i, \sigma_j \, = \, \uparrow \uparrow \text{ or } \uparrow \downarrow$, respectively.
We employ the 6-31G basis set \cite{hehreSelfConsistentMolecular1972}
for the initial RHF and PM calculations, and then add nuclear cusps to the
AOs using the \textsc{cgaows} package
\cite{quadyMethodindependentCuspsAtomic2024,quadyCuspingGaussianAtomicOrbitalsWithSlaters2024} 
before starting our all-electron VMC calculations.

All geometries for the alkene molecules
were optimized at the MP2/aug-cc-pVTZ level of theory
and are available in the SI.
The Jastrow factor parameter $A$ was chosen via an initial
minimization with the RHF orbitals held fixed, resulting in the
value $A=0.0225$ for propene and butene and $A=0.0200$ for pentadiene.
All VMC calculations from this work were computed
using our own all-electron VMC software and LM implementation. 
At each LM iteration, the diagonal shift $a$ was initially set to 0.01.
To exert a basic form of step size control,
we inspected the energy change predicted by Eq.\ (\ref{eq:Hdp=eSdp})
and, if no $\lambda$ lowered the energy within 0.1 to 0.0 $\mathrm{E_h}$ or resulted in any LCAO coefficient change of more than 0.25,
we increased $a$ by a factor of ten and recalculated the update
(repeating as necessary until the energy change and maximum parameter change were below their respective thresholds).
For the first 18 iterations of the LM, we employed our expand-and-prune
sLCAO algorithm, after which the variational parameter set was locked in and 22 subsequent LM steps were made using the standard LM procedure.
The VMC sample sizes were set to
$1\,971\,200$, $2\, 628\, 192$, and $3\, 120\, 992$ 
in propene, butene, and pentadiene, respectively.
The rising sample size across this series was used to counteract
the growth in the variance as the number of electrons increased.
To ensure that the stochasticity of the optimization itself was accounted for
in our reported results, we repeated each optimization with five different
random seeds.
We took the lowest 10 energies from the last 20 iterations of each optimization
(50 energies overall) and evaluated their average and standard error,
the latter via a blocking analysis \cite{flyvbjergErrorEstimatesAverages1989}
as shown in the SI.

\begin{figure*}[]
    \centering
    \includegraphics[width=0.95\linewidth]{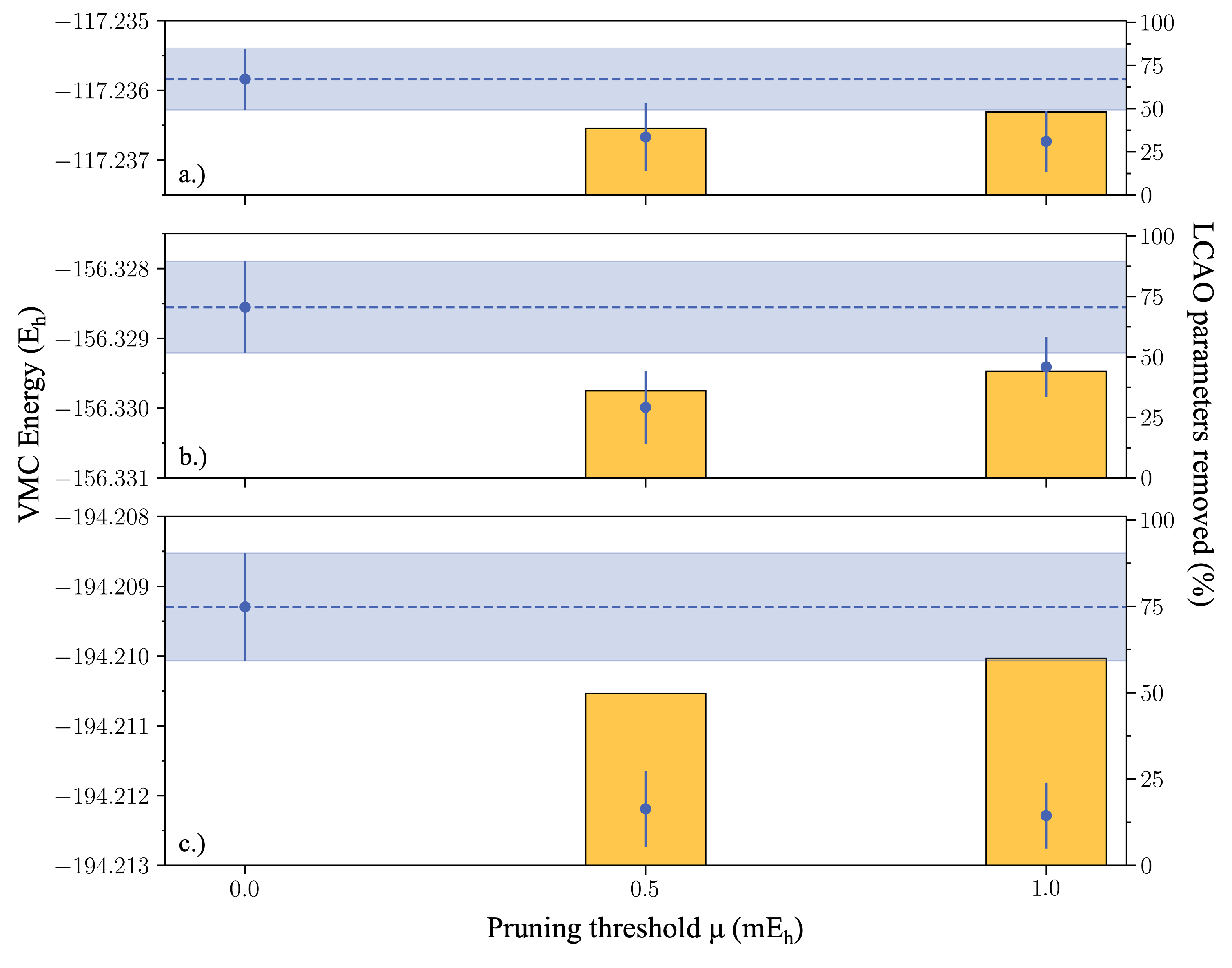}
    \caption{
    Optimized energies (left axis and points) and the fraction
    of LCAO parameters pruned away (right axis and orange bars)
    for standard LM ($\mu=0$) and sLCAO optimizations in
    a.) propene, b.) butene, and  c.) pentadiene.
    Blue shading and dashed lines are guides to the eye.
    }
    \label{fig:AbsE_v_ep}
\end{figure*}

\section{Results}
\label{sec:results}

The efficacy of the sLCAO approach compared to the standard LM
in three alkene molecules is shown in Fig.\ \ref{fig:AbsE_v_ep}.
Looking at the energy (left) axis, we see that sLCAO with either
$\mu=1.0~\mathrm{mE_h}$ or $\mu=0.5~\mathrm{mE_h}$ produces energies
that are at least as low as those from the standard LM ($\mu=0$).
In three cases, they are lower by a statistically significant amount.
Indeed, in pentadiene, the largest molecule where we would expect
sLCAO to make the biggest difference, sLCAO improves on the standard LM
by about 3 $\mathrm{mE_h}$ using either of the two thresholds we tested.
We should stress that the sLCAO optimizations used exactly the same
sample size as the standard LM optimizations, and both used the same
step size control.

Looking at the right axis of Fig.\ \ref{fig:AbsE_v_ep}, we see that
the sLCAO method generally prunes away a increasingly large fraction of the
LCAO parameters as the molecule size increases from propene to
butene to pentadiene.
This behavior is what we would expect, since, in the limit of a large
molecule, a localized MO should only have significant contributions
coming from $O(1)$ AOs.
By the time we get to pentadiene, which, it must be said, is not
that large of a molecule, the method is pruning away 50\% or more
of the LCAO parameters while delivering improved variational outcomes.
This result appears to confirm that many of the LCAO parameters are,
at least at the sample sizes we are using, small to the point that
their only role in a standard LM optimization is to introduce additional
noise that degrades the optimization's effectiveness for the more
significant parameters.

\begin{figure}[]
    \centering
    \includegraphics[width=0.95\linewidth]{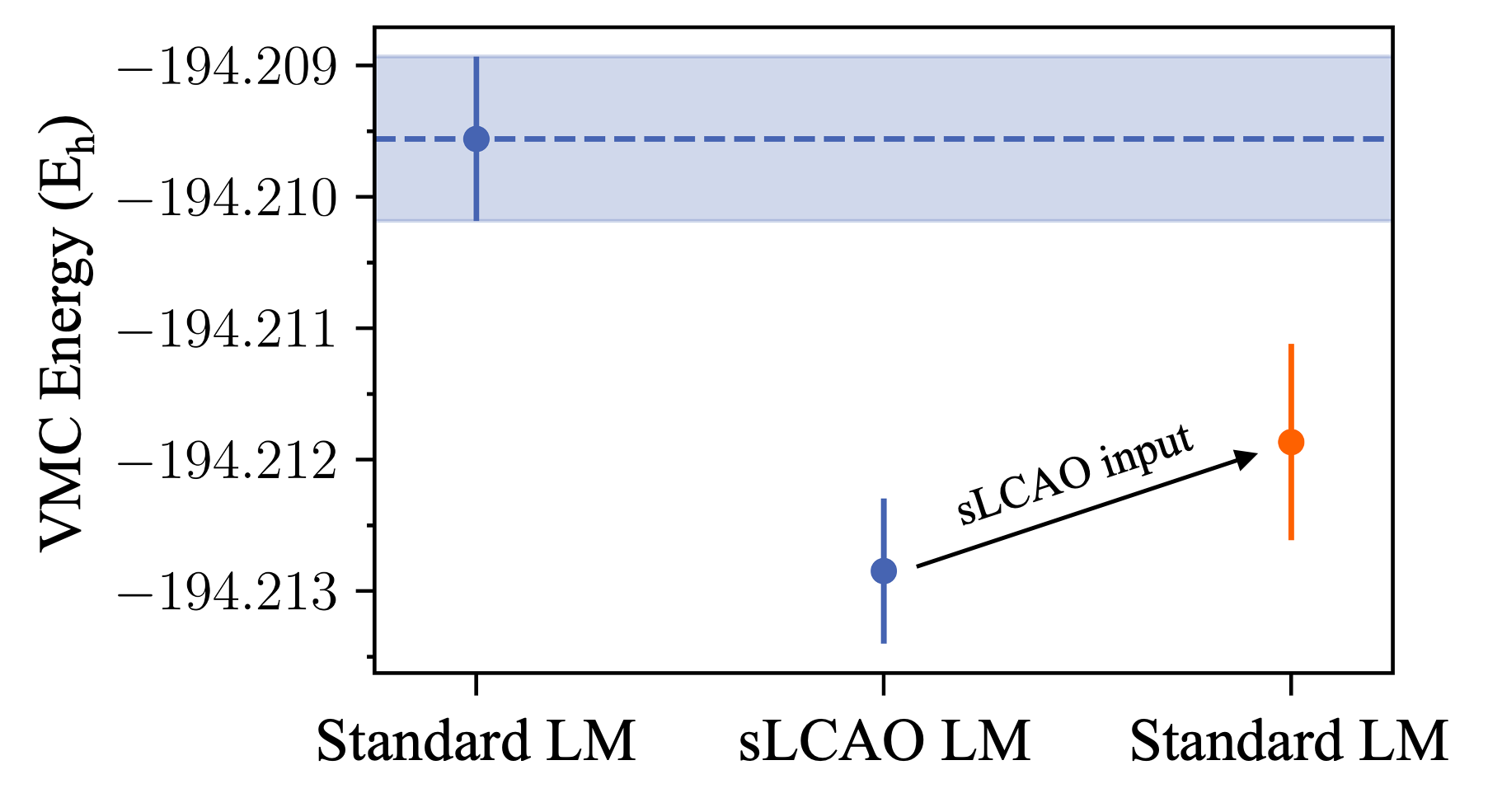}
    \caption{
    Comparison of optimized energies for a single standard LM optimization starting from RHF (left),
    sLCAO with $\mu=0.5~\mathrm{mE_h}$ (center), and the standard LM started
    from the sLCAO result (right) in pentadiene.
    Shading and dotted line are guides to the eye.
    }
    \label{fig:sLCAO2Vanilla}
\end{figure} 


Seeing an ansatz with a reduced parameter set achieve an improved
variational outcome is not the norm, and so we think it is important
to verify the quality of the resulting wave function.
To do so, we perform a test across single LM optimizations (VMC result now averaged from 10 energies) in which the final parameters from one of the
$\mu=0.5~\mathrm{mE_h}$ sLCAO optimizations of pentadiene are used
as the initial guess in a subsequent standard LM optimization.
In Fig.\ \ref{fig:sLCAO2Vanilla}, we see that this LM optimization
reproduces the sLCAO energy (at least within statistical uncertainty)
and confirms that we have found a variationally superior wave function
as compared to what was found by the standard LM optimization that
started from localized RHF.
If anything, the additional noise present in the subsequent LM optimization
appears to have slightly degraded the quality of the wave function
compared to the sLCAO result, although it is possible that the rise
in energy is simply statistical scatter.
We again conclude that, in the context of a stochastic second order
optimization in which one parameter's uncertainty can impact the quality
of another parameter's update, it is profitable to construct a selection
mechanism by which unimportant parameters are switched off entirely,
as the small improvement they make in formal variational flexibility
is not worth the optimization step noise they produce in practice.

\section{\label{sec:conclusion}Conclusion}

We tested an algorithm for selectively disabling the least important
LCAO coefficients during a linear method orbital optimization and found that
doing so improves the optimization.
Simple tests in collections of H$_2$ molecules clarified that
the presence of unimportant LCAO parameters, even when initialized to
zero, substantially degraded the statistical quality of the
linear method update for the important parameters.
To counteract this effect, we designed a Fock-operator-based,
single-threshold pruning algorithm that removes energetically unimportant
LCAO parameters from the optimization on the fly.
We coupled this pruning with a simple orbital expansion step that
re-enables the parameters at the edges of each local orbital, producing
an overall expand-and-prune sLCAO method in which the optimization
itself discovers the balance point between including and excluding parameters.
In tests on propene, butene, and pentadiene, we found that the sLCAO
approach improves on the standard LM, producing lower energies when
given the same starting point and sample size.

\section*{Supplementary Information}
See \href{run:SI.tex}{Supplementary Information} for geometries of the hydrogen toy systems and alkenes, along with details of the blocking standard error analysis.

\begin{acknowledgments}
This work was supported by the Office of Science, Office of Basic Energy Sciences, 
the U.S. Department of Energy, Contract Number DE-AC02-05CH11231, 
through the Gas Phase Chemical Physics program. Computational work was performed 
with the LBNL Lawrencium cluster and the Savio computational cluster resource 
provided by the Berkeley Research Computing program at the University of 
California, Berkeley. T.K.Q. acknowledges that this material is based upon work 
supported by the National Science Foundation Graduate Research Fellowship Program 
under Grant No. DGE 2146752. Any opinions, findings, and conclusions or recommendations 
expressed in this material are those of the authors and do not necessarily reflect 
the views of the National Science Foundation.

\end{acknowledgments}

\section*{Data Availability Statement}

The data that support the findings of this study are available
within the article and its supplementary material.

\section*{References}

\bibliographystyle{achemso}
\bibliography{references}

\clearpage
\onecolumngrid
\section{Supplementary Information}
\renewcommand{\thesection}{S\arabic{section}}
\renewcommand{\theequation}{S\arabic{equation}}
\renewcommand{\thefigure}{S\arabic{figure}}
\renewcommand{\thetable}{S\arabic{table}}
\setcounter{section}{0}
\setcounter{figure}{0}
\setcounter{equation}{0}
\setcounter{table}{0}
\section{Geometries}
All coordinates are in Bohr.
\subsection{H\textsubscript{2}}
\vspace{-0.75cm}
\begin{table}[h]
\centering
\begin{tabular}{crrr} 
\hline \hline
Z & x & y & z \\
\hline
H &  -0.384022   &      0.339143   &     -0.467775 \\  
H &   0.384022   &      1.080041   &      0.467775 \\
\hline
\hline \hline
\end{tabular}
\end{table}
\subsection{(H\textsubscript{2})\textsubscript{4}}
\vspace{-0.75cm}
\begin{table}[h]
\parbox{.45\linewidth}{
\centering
\begin{tabular}{crrr} 
\hline \hline
Z & x & y & z \\
\hline
H &    0.434635 &  0.183906 &  5.864799 \\  
H &   -0.434635 &  1.235278 &  5.473554 \\
H &    0.217190 &  0.057499 & 11.514774 \\
H &   -0.217190 &  1.361686 & 11.161936 \\
H &    0.242382 &  6.951057 &  5.326739 \\
H &   -0.242382 &  5.806484 &  6.011610 \\
H &    0.302443 &  6.860864 & 11.762185 \\
H &   -0.302443 &  5.896677 & 10.914522 \\
\hline \hline
\end{tabular}
}
\end{table}
\subsection{(H\textsubscript{2})\textsubscript{9}}
\vspace{-0.75cm}
\begin{table}[h]
\centering
\begin{tabular}{crrr} 
\hline \hline
Z & x & y & z \\
\hline
H  &  0.434635 &   0.183906 &   5.864799  \\   
H  & -0.434635 &   1.235278 &   5.473554  \\
H  &  0.217190 &   0.057499 &  11.514774  \\
H  & -0.217190 &   1.361686 &  11.161936  \\
H  & -0.384023 &   0.339143 &  16.665094  \\
H  &  0.384023 &   1.080041 &  17.349967  \\
H  &  0.242382 &   6.951057 &   5.326739  \\
H  & -0.242382 &   5.806484 &   6.011610  \\
H  &  0.302443 &   6.860864 &  11.762185  \\
H  & -0.302443 &   5.896677 &  10.914522  \\
H  &  0.434635 &   5.853085 &  17.203156  \\
H  & -0.434635 &   6.904457 &  16.811911  \\
H  &  0.217190 &  -5.726677 &   5.845596  \\
H  & -0.217190 &  -7.030864 &   5.492757  \\
H  &  0.242382 &  -6.951057 &  10.995916  \\
H  & -0.242382 &  -5.806484 &  11.680789  \\
H  &  0.434635 &  -5.853085 &  17.203156  \\
H  & -0.434635 &  -6.904457 &  16.811909  \\
\hline \hline
\end{tabular}
\end{table}

\newpage

\subsection{Propene}
\vspace{-0.75cm}
\begin{table}[h]
\centering
\caption{Equilibrium geometry.
} 
\begin{tabular}{crrr} 
\hline \hline
Z & x & y & z \\
\hline
C &   2.320652  &  -0.309570  &  -0.000000 \\ 
C &  -0.246422  &   0.871176  &  -0.000000 \\
C &  -2.412108  &  -0.422579  &   0.000000 \\
H &  -2.417997  &  -2.468600  &  -0.000000 \\
H &  -4.218831  &   0.529655  &   0.000000 \\
H &  -0.324157  &   2.920360  &  -0.000000 \\ 
H &   2.188048  &  -2.363489  &  -0.000000 \\ 
H &   3.399481  &   0.273346  &  -1.657376 \\
H &   3.3994810  &   0.273346  &   1.657376 \\
\hline \hline
\end{tabular}
\end{table}
\subsection{Butene}
\vspace{-0.75cm}
\begin{table}[h]
\parbox{.45\linewidth}{
\centering
\caption{Equilibrium geometry.
} 
\begin{tabular}{crrr} 
\hline \hline
Z & x & y & z \\
\hline
C    &    1.012654  &  -0.754809  &   0.000000 \\
C    &   -1.012655  &   0.754810  &   0.000001 \\
C    &    3.690136  &   0.152640  &  -0.000000 \\
C    &   -3.690136  &  -0.152640  &  -0.000000 \\
H    &    0.713619  &  -2.786947  &   0.000000 \\
H    &   -0.713620  &   2.786948  &   0.000001 \\
H    &    3.769973  &   2.209740  &   0.000001 \\
H    &   -3.769971  &  -2.209741  &   0.000001 \\
H    &    4.706222  &  -0.533928  &   1.657785 \\
H    &    4.706221  &  -0.533927  &  -1.657787 \\
H    &   -4.706222  &   0.533927  &   1.657785 \\
H    &   -4.706221  &   0.533926  &  -1.657787 \\
\hline \hline
\end{tabular}
}
\end{table}
\subsection{Pentadiene}
\vspace{-0.75cm}
\begin{table}[h]
\centering
\caption{Equilibrium geometry.
} 
\begin{tabular}{crrr} 
\hline \hline
Z & x & y & z \\
\hline
C  &    4.759477  &  -0.496575  &  -0.000000  \\
C  &    2.555821  &   0.758238  &   0.000000  \\ 
C  &    0.100453  &  -0.462993  &   0.000001  \\
C  &   -2.100388  &   0.801668  &   0.000000  \\
C  &   -4.646587  &  -0.414893  &  -0.000000  \\
H  &    4.803250  &  -2.541997  &  -0.000001  \\
H  &    6.545736  &   0.492136  &  -0.000001  \\
H  &    2.568129  &   2.809810  &   0.000001  \\
H  &    0.074766  &  -2.516830  &   0.000001  \\
H  &   -2.039382  &   2.854299  &  -0.000001  \\
H  &   -4.485731  &  -2.466724  &  -0.000000  \\
H  &   -5.736261  &   0.151316  &   1.657203  \\
H  &   -5.736260  &   0.151316  &  -1.657203  \\
\hline \hline
\end{tabular}
\end{table}

\newpage
\section{Blocking analysis for standard error}\label{SIsec:blocking}
Blocking analysis to determine the standard error of the mean energy from Flyvbjerg.~\cite{flyvbjergErrorEstimatesAverages1989} 
To remove the correlation from the set of energies $\{x_i\}$ of length $N_T$, calculated by VMC, first block the data in to $N_b$ equipartitioned chunks of length $L_b = N_T / N_b$. 
Average the contents from each block creating a new set $\{m_k\}$ of length $N_b$.  

\begin{align}
    \overline{x} &= \frac{1}{N_T} \sum_{i=0}^{N_T} x_i\\
    m_j &= \frac{1}{L_b} \sum_{i=(j-1)L_b+1}^{j L_b} x_i\\
\end{align} 

The standard error of the mean energy ($\overline{x}$) is the square root of the variance of the mean ($\sigma^2_{\overline{x}} (m)$) where $N_b$ is chosen when standard error plateaus with respect to $N_b$. $\sigma^2_{\overline{x}} (m)$ is not to be confused with the variance of the block $c_0$ (the spread).

\begin{align}
    c_0 &= \frac{1}{N_b} \sum_{j=1}^{N_b} (m_j - \overline{x})^2 \\
    \sigma^2_{\overline{x}} (m) &= \frac{c_0}{N_b-1} \\
    \text{Standard error} \equiv \sigma(m) &= \sqrt{\sigma^2_{\overline{x}}(m)}
\end{align}

\end{document}